\documentclass[
reprint,
bibnotes,
amsmath,amssymb,
aps,
]{revtex4-1}

\usepackage{graphicx}
\usepackage{dcolumn}
\usepackage{bm}
\usepackage{hyperref}
\usepackage{csquotes}
\hypersetup{colorlinks=true,linkcolor=red,citecolor=red}
\usepackage{float}

\begin{document}

\title{Interplay between chemical order and magnetic properties in L1$_0$ FeNi (tetrataenite): A First-Principles Study}

\author{Ankit Izardar}
\author{Claude Ederer}
\affiliation{Materials Theory, ETH Z\"urich, Wolfgang-Pauli-Strasse 27, 8093 Z\"urich, Switzerland}

\date{\today}

\begin{abstract}

We use first-principles-based calculations to investigate the interplay between chemical order and the magnetic properties of $L1_0$ FeNi. In particular, we investigate how deviations from perfect chemical order affect the energy difference between the paramagnetic and ferromagnetic states as well as the important magneto-crystalline anisotropy energy. Our calculations demonstrate a strong effect of the magnetic order on the chemical order-disorder transition temperature, and conversely, a strong enhancement of the magnetic transition temperature by the chemical order.
Most interestingly, our results indicate that the magnetic anisotropy does not decrease significantly as long as the deviations from perfect order are not too large. Moreover, we find that in certain cases a slight disorder can  result in a higher anisotropy than for the fully ordered structure. We further analyze the correlation between the magneto-crystalline anisotropy and the orbital magnetic moment anisotropy, which allows to study the effect of the local chemical environment on both quantities, potentially enabling further optimization of the magneto-crystalline anisotropy with respect to chemical order and stoichiometric composition.
\end{abstract}

\maketitle

\section{\label{sec:Intro}Introduction}

Magnetic materials are ubiquitous and play a pivotal role in many technological applications ranging from consumer electronic devices to electric power production and conversion.
In particular, high performance permanent magnets form crucial components in the devices used for generating electric power from renewable energy sources such as wind, hydro, tidal, etc. 
The strength of a permanent magnet is quantified by the maximum magnetic energy product $(BH)_\text{max}$, i.e, the product of the remanence $B_r$ and the coercivity $H_c$. 
Thus, high performance permanent magnets are typically composed of rare-earth elements (Sm, Nd, Dy, etc.), which provide high resistance to demagnetization, in combination with transition-metals (Fe, Co, etc.), which provide high saturation magnetization.
Specifically, magnets belonging to the SmCo family (e.g. SmCo$_5$ and Sm$_2$Co$_{17}$), with energy products in the range of 5-20 MGOe (40-160 kJ/m$^3$) \cite{Strnat1967,Hoffer1967}, and the NdFeB family (e.g. Nd$_2$Fe$_{14}$B), with energy products in the range 5-50 MGOe (40-400 kJ/m$^3$) \cite{Herbst1991}, are currently the best-performing \emph{supermagnets}.
However, the volatility in price and uncertainty of supply of the required rare earth elements, makes it highly desirable to find alternatives to these rare-earth based magnets, in order to meet the increasing global demand for permanent magnets~\cite{SMITHSTEGEN20151, McCallum2014}.

\begin{figure*}
    \includegraphics[width=0.9\textwidth]{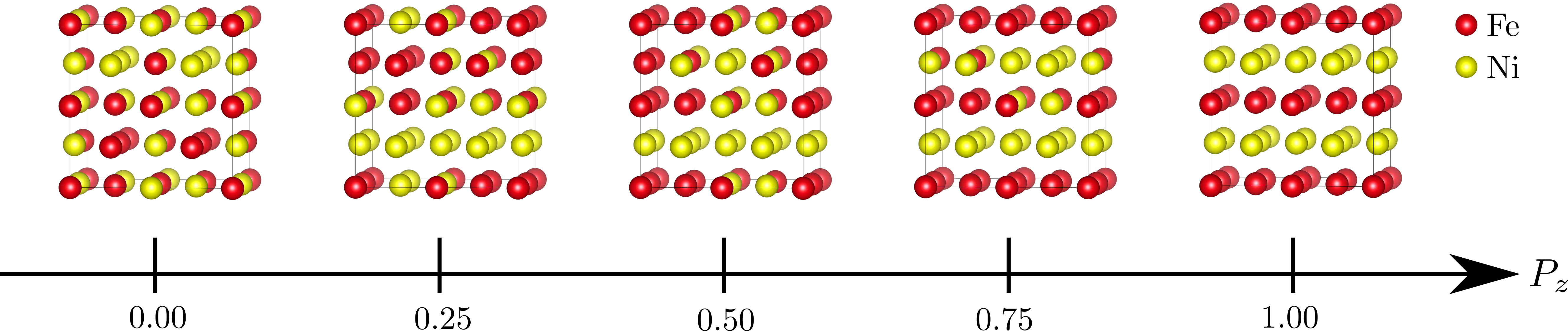}
    \caption{Examples of atomic configurations corresponding to different values of the long range order parameter $P_z$, depicted in a $2\times2\times2$ supercell relative to the conventional 4-atom cubic cell. Fe and Ni atoms are represented by red and yellow spheres, respectively. In the chemically ordered L1$_0$-FeNi phase (tetrataenite) with $P_z=1$, Fe and Ni atoms occupy alternate layers perpendicular to $c$. In the fully disordered A1 phase ($P_z=0$), they randomly occupy sites of the underlying fcc lattice.}
    \label{img:FeNi}
\end{figure*}

An interesting candidate in this respect is the chemically-ordered L1$_0$ phase of Fe$_{50}$Ni$_{50}$ (tetrataenite), which has been found in iron meteorites~\cite{ALBERTSEN1978, PETERSEN1977192, Danon1979, Danon1980, clarke}.
The Fe and Ni atoms in tetrataenite occupy alternating planes of the underlying fcc lattice oriented perpendicular to the $c$ axis (see rightmost graph in Fig.~\ref{img:FeNi}), resulting in a structure with tetragonal symmetry and a high magneto-crystalline anisotropy energy (MAE) ($> 7 \cdot 10^6$\,erg\,cm$^{-3}$), large saturation magnetization ($\sim 1270$\,emu\,cm$^{-3}$), and a projected energy product of 42\,MGOe (335\,KJ/m$^3$)  \cite{doi:10.1063/1.1656361,Kojima2014,Lewis2014InspiredBN,Lewis2014}.
In contrast, the disordered phase, where Fe and Ni atoms are randomly distributed over the sites of the fcc lattice (see leftmost graph in  Fig.~\ref{img:FeNi}), exhibits only a very small MAE.
 
Unfortunately, the laboratory synthesis of the ordered phase is extremely challenging due its rather low order-disorder transition temperature, $T_\text{od} \sim$ 593\,K~\cite{neel}, and the slow diffusion of atoms at this temperature, which is of the order of one atomic jump per $10^{4}$ years at 573\,K \cite{Scorzelli1997}. Since its discovery, several attempts have been made to achieve a high degree of chemical order in this alloy \cite{Takata_2018, SHIMA20072213, Makino2015, Goto2017}. Nevertheless, synthesis of a fully ordered system remains challenging.

The low order-disorder temperature and the difficulties in synthesizing fully ordered samples make it also very challenging to fully characterize the magnetic properties of tetrataenite, as the disordering occurs below the predicted Curie temperature. It also raises the question of how the favorable magnetic properties depend on the degree of chemical order.

Several previous studies have found a strong coupling between the magnetic and chemical orders in this system.
For example, both Dang \textit{et al.} \cite{Dang1996} and Lavrentiev \textit{et al.} \cite{Lavrentiev2014} found, using different models and approximations, that the ferromagnetic Curie temperature is drastically enhanced in the chemically ordered case compared to the random alloy (from $\sim$ 450\,K to over 1000\,K in Ref.~\onlinecite{Lavrentiev2014}) and that also the magnetic interactions strongly increase the chemical order-disorder transition temperature (by $\sim 100$\,K in Ref.~\onlinecite{Dang1996}).
This suggests that it is necessary to include both chemical and magnetic degrees of freedom to accurately describe this system.

In this work, we present additional complementary insights on the interplay between chemical order and magnetic properties in tetrataenite by means of first-principles-based density functional theory (DFT) and Monte Carlo simulations. In particular, we study how the MAE depends on the degree of chemical order in the system. We find that small deviations from perfect order do not lead to a significant reduction of the magnetic anisotropy, and that in some cases a small amount of disorder can even enhance the MAE.
We then discuss the anisotropy of the local orbital moments as an indicator that allows to further optimize the magnetic anisotropy with respect to the local atomic environment.

The remainder of the paper is structured as follows. In Sec.~\ref{sec:methods} we first define the long range order parameter, then describe how we model the partially disordered as well as the paramagnetic state in FeNi, and introduce the computational methods used throughout this work. In Sec.~\ref{sec:results}, we then discuss our results regarding the energetics of the order-disorder transition, the effect of chemical disorder on the MAE, and the correlation between orbital magnetic moment anisotropy and the MAE. Finally, in Sec. \ref{sec:Summary}, we conclude by summarizing our main findings.      

\section{Models and methods}
\label{sec:methods}

\subsection{\label{sec:order-disorder}Modeling of chemical disorder}

\begin{figure}
   \centering
   \includegraphics[width=0.5\columnwidth]{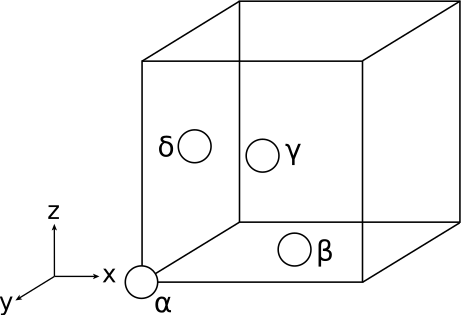}
   \caption{Depiction of the four sites of the fcc lattice within the conventional cubic unit cell, defining the four sublattices $\alpha$, $\beta$, $\gamma$, and $\delta$.}
   \label{img:sublattice}
\end{figure}

To define the long range order parameter for the $L1_0$ chemical order, we divide the fcc lattice into four individual sublattices, $\alpha$, $\beta$, $\gamma$, and $\delta$, according to the four different sites in the conventional 4-atom cubic unit cell (see Fig.~\ref{img:sublattice}).
The fully ordered $L1_0$ structure can then be described in three different ways, corresponding to arrangements of different atomic species in alternating planes perpendicular to the three Cartesian axes.
Thereby, always two sublattices are fully occupied by one type of atom, while the other two sublattices are occupied by the other type.
For example, alternating atomic planes perpendicular to $z$ correspond to occupation of sublattice $\alpha$ and $\beta$ by one type of atom and occupation of sublattices $\gamma$ and $\delta$ by the other type, whereas for alternating planes perpendicular to $x$, sublattices $\alpha$ and $\delta$ are occupied by one type of atom and sublattices $\beta$ and $\gamma$ by the other type.

We can now define long range order parameters for the three different orientations of the $L1_0$ order as follows:
\begin{equation}\label{eq:px}
P_x  = p_{\alpha}^\text{Fe} + p_{\gamma}^\text{Fe} - 1 \quad ,
\end{equation}
\begin{equation}\label{eq:py}
P_y  = p_{\alpha}^\text{Fe} + p_{\delta}^\text{Fe} - 1 \quad ,
\end{equation}
\begin{equation}\label{eq:pz}
P_z  = p_{\alpha}^\text{Fe} + p_{\beta}^\text{Fe} - 1 \quad ,
\end{equation}
where $p_i^\text{Fe}$ is the probability that a site on sublattice $i$ is occupied by an Fe atom.
These probabilities have to fulfill the condition $\sum_i p_i^\text{Fe} = 2$ (on average 2 Fe atoms per 4-atom unit cell), and thus only three can be chosen independently. Furthermore, each $p_i^\text{Fe}$ can only vary between 0 and 1, imposing an additional constraint on the $p_i^\text{Fe}$.
Nevertheless, Eqs.~\eqref{eq:px}-\eqref{eq:pz} can be inverted and the probabilities $p_i^\text{Fe}$ are then uniquely defined by specifying the three components of the long range order parameter within the allowed range. 

To model the system with a given value for the long-range order parameter, we generate 50 configurations, using a $2\times2\times2$ supercell of the conventional cubic cell. For each configuration, we randomly distribute 16 Fe and 16 Ni atoms over the 32 available sites, according to the probabilities $p_i^\text{Fe}$ corresponding to a fixed value of $P_z$ and $P_x=P_y=0$.
The chosen supercell size allows to obtain five different values for the long range order parameter, $P_z \in \{ 0, 0.25, 0.5, 0.75, 1\}$. 
We then calculate the total energy for each configurations using density functional theory (DFT), as described in Sec.~\ref{sec:computational_details}. The total energy for a given order parameter is then obtained by averaging over the corresponding configurations.

We note that most previous first-principles-based studies, e.g., Ref.~\onlinecite{Tian2019}, have employed effective medium/mean-field type approaches to model the compositional disorder. While our complementary approach is computationally more demanding, since it requires sampling over many configurations, it also incorporates effects of disorder within the local environment, which turns out to be especially important in the case of the MAE.

\subsection{\label{sec:DLM}Modeling of the paramagnetic state}

It is well known that in most magnetic materials local magnetic moments still exist above the Curie temperature, even though the material does not exhibit any macroscopic (long-range) magnetic order. The incorporation of such local moments is very important to correctly describe the electronic structure of these materials, and thus the paramagnetic phase cannot simply be treated as a non-magnetic state in DFT-based first-principles calculations.

In order to model the paramagnetic state, we therefore employ the disordered local moment (DLM) method~\cite{STAUNTON198415}, where the directions of magnetic moments are constrained to random directions. Analogously to our treatment of chemical disorder, we use a supercell approach and sample over a sufficient amount of randomly generated configurations~\cite{Alling2010}. 
The average of the energy over all configurations then represents the energy of the paramagnetic phase (in the limit of very high temperature). 

For the chemically ordered case, we generate 100 collinear DLM configurations by randomly initializing the magnetic moments of the Fe atoms in a $2 \times 2 \times 2$ supercell as either up or down. For the chemically disordered case, we create 10 different chemically disordered configurations (as described in Sec.~\ref{sec:order-disorder}) and then generate 10 DLM configurations for each of these configurations. 
We do not explicitly initialize the Ni magnetic moments, since the Ni moments tend to vanish if the surrounding Fe magnetic moments are oriented anti-parallel to each other. In other cases, the Ni moments will converge to either up or down, depending on the orientation of moments on the surrounding Fe atoms. 
Therefore, we do not take into account the directions of the Ni moments as independent variables.
We also do not consider any noncollinear configurations. These are not expected to alter the results if the basic assumptions of the DLM method are valid, but would significantly increase the required computational effort.

To verify our sampling of the paramagnetic state, we evaluate the nearest-neighbor spin-correlation function for the magnetic moments of the Fe atoms~\footnote{In analogy to the creation of our DLM configurations, we consider only the Fe magnetic moments when evaluating the spin-correlation function.}:
\begin{equation}
\label{eq:spincorr}
\Phi  = \frac{1}{N_\text{Fe}} \sum_{i} \frac{1}{N_{i}} \sum_j \hat{e_i} \cdot \hat{e}_j \quad , 
\end{equation}
where the sum over $i$ goes over all $N_\text{Fe}$ Fe atoms in the supercell ($N_\text{Fe}=16$ in the present case), the sum over $j$ goes over all Fe nearest neighbors for each $i$ (with $N_i$ being the number of Fe nearest neighbors of atom $i$, which is different for each individual configuration), and $\hat{e}_i$ is the direction of the magnetic moment of Fe atom $i$.

\subsection{\label{sec:computational_details}Computational methods}

All DFT calculations are performed using the Vienna \textit{ab} \textit{initio} Simulation package (VASP) \cite{Kresse1996}, the projector-augmented wave method (PAW) \cite{PAW1994,Kresse1999}, and the generalized gradient approximation according to Perdew, Burke, and Ernzerhof~\cite{PBE}. Brillouin zone integrations are performed using the tetrahedron method with Bl\"{o}chl corrections and a $\Gamma$-centered $14\times14\times14$ \textbf{k}-point mesh. The plane wave energy cut-off is set to 350\,eV, and the total energy is converged to an accuracy of $10^{-8}$\,eV. 
Our PAW potentials include 3$\textit{p}$, 4$\textit{s}$, and 3$\textit{d}$ states in the valence for both Fe and Ni .   

The MAE is calculated using the magnetic force theorem \cite{LIECHTENSTEIN198765,Daalderop1991}, i.e., by including the spin-orbit coupling in a non-self-consistent calculation, using the charge density converged without spin-orbit coupling, and then taking the difference in energies between two different orientations of the magnetization direction.

We define the MAE as the energy difference $E^{[100]} - E^{[001]}$, where $E^{[100]}$ and $E^{[001]}$ are the total energies obtained with magnetization aligned along the $[100]$ and $[001]$ directions, respectively. Thus, the MAE is defined as positive when the magnetic easy axis lies along the $[001]$ direction, which is the reported easy axis for $L1_0$ FeNi~\cite{neel, Lewis2014}. To check the convergence of the MAE with respect to the \textbf{k}-point sampling, we perform calculations using up to $25\times25\times25$ \textbf{k}-points and find that the MAE is sufficiently converged (to about $\pm 1\,\mu$eV/f.u.) for our purposes using a $14\times14\times14$ \textbf{k}-point mesh.

The temperature dependence of the chemical long-range order parameter is obtained from simple Monte Carlo simulations, considering an fcc lattice using a $\sqrt[3]{N} \times \sqrt[3]{N} \times \sqrt[3]{N}$ supercell of the conventional cubic cell, containing $4N$ sites over which we distribute Fe and Ni atoms in equal proportion. For a given temperature, we perform Monte Carlo sweeps using the Metropolis algorithm, where in each trial step the configuration is varied by exchanging the positions of an arbitrarily chosen pair of Fe and Ni atoms, then calculating the long range order parameter $\mathbf{P} = (P_x, P_y, P_z)$, and evaluating the corresponding total energy as described in Sec.~\ref{subsec:ferro}.

\section{\label{sec:results}Results and Discussion}

\subsection{\label{subsec:ferro} Energetics of the order-disorder transition}

We first determine equilibrium lattice parameters for perfectly ordered $L1_0$ FeNi in the ferromagnetic state. 
We obtain lattice parameters $a = 3.560$\,{\AA} and $c = 3.577$\,{\AA} ($c/a = 1.0048$). Our calculated lattice parameters agree well with the values measured in experiments ($a = 3.560$\,{\AA} to $3.582$\,{\AA} and $c = 3.589$\,{\AA} to $3.615$\,{\AA})~\cite{Kotsugi_2014, Makino2015}, and obtained in previous calculations ($a = 3.557$\,{\AA} to $3.560$\,{\AA} and $c = 3.570$\,{\AA} to $3.584$\,{\AA})~\cite{Lewis2014InspiredBN, Alex,Miura2013}.

Next, we determine the dependence of the total energy on the long range chemical order parameter, while keeping the perfect ferromagnetic order. For this, we calculate the total energy  of 50 configurations for each value of $P_z$, generated as described in Sec.~\ref{sec:order-disorder}.

For simplicity, we keep the lattice parameters fixed corresponding to a metrically cubic unit cell with $a$ = 3.560 \AA{} and $c/a$ = 1, i.e., we neglect the small tetragonal strain on the unit cell (which will also depend on the degree of long range order).
Our test calculations for perfect chemical order ($P_z = 1$) show that these simplifications change the total energy by less than 5 meV/atom, which is negligible compared to the energy changes related to the different distributions of atoms.
Furthermore, we do not perform any further optimization of atomic coordinates for the disordered configurations.

\begin{figure}[t]
   \centering
   \includegraphics[width=\columnwidth]{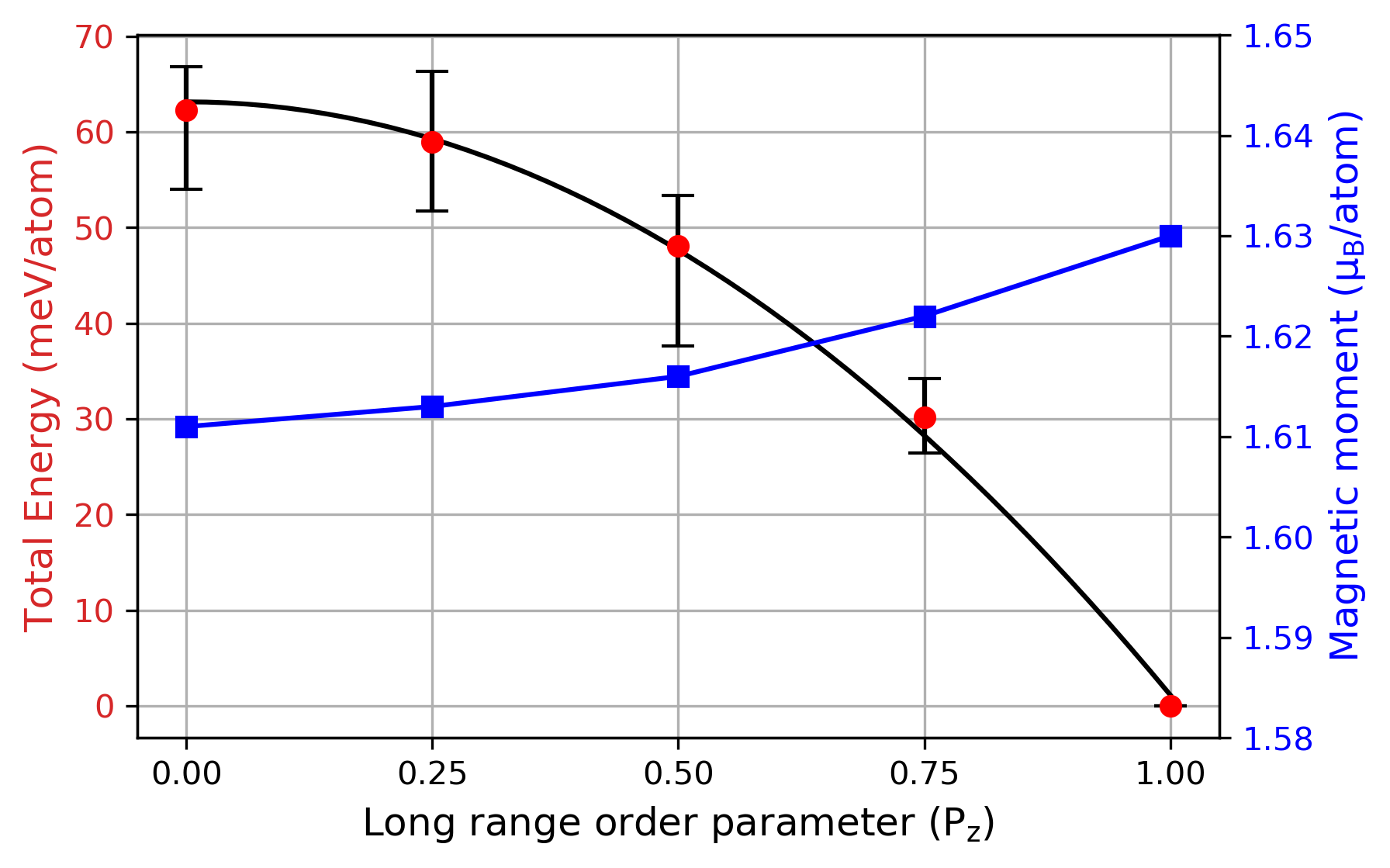}
   \caption{Total energy (per atom) and average magnetic moment (per atom) as function of the long range order parameter $P_z$ (with $P_x=P_y=0$), calculated for perfect ferromagnetic order. Red dots represent the mean over 50 configurations for each \textit{P$_z<1$}. Error bars indicate the highest and lowest energies of the individual configurations. The black curve shows a quadratic fit to the data. Energies are defined relative to the fully ordered state ($P_z=1$). Filled squares represent the mean of magnetic moments over 50 configurations for each \textit{P$_z<1$}. The blue line is a guide to the eye.}
   \label{img:e_vs_p}
\end{figure}

The corresponding total energies (averages as well as total spread over different configurations) are shown in Fig.~\ref{img:e_vs_p} as function of the long range order parameter $P_z$, together with the average magnetic moment per atom.
It can be seen that the averaged total energies are well fitted by a quadratic dependence on $P_z$, $E = E_0 - \Delta E P_z^2$, where $\Delta E = 62$\,meV is the energy difference (per atom) between the perfectly ordered and completely disordered structure. 
We note that a quadratic dependence on $P_z$, or more generally on $P=\sqrt{P_x^2+P_y^2+P_z^2}$ also corresponds to the leading order term allowed by symmetry for small fluctuations around the disordered state, $P=0$, and also follows from a simple energetic model with only nearest neighbor interactions.
The good quality of the quadratic fit thus also indicates that rather accurate (sufficient for our purposes) mean energies can be obtained by using 50 different configurations for each $P_z<1$.

One can also see that the average total magnetic moment depends only weakly on $P_z$, increasing slightly from 1.611 $\mu_B$ to 1.630 $\mu_B$ between zero and full chemical order. We note that the increase in the total magnetic moment is mainly due to the average magnetic moment of the Fe atoms, while the average Ni magnetic moment remains fairly constant until $P_z = 0.75$, after which it slightly decreases for the perfectly ordered structure. 

These results agree very well, both qualitatively and quantitatively, with recent calculations by 
Tian \textit{et al.} employing the coherent potential approximation (CPA) to treat the compositional disorder~\cite{Tian2019}. The good agreement between this complementary approach and our configurational sampling technique confirms on one side the good convergence of our data and on the other side also indicates that effects of the local environment, not included in the CPA approach, are not too relevant for the total energy and average magnetic moment.

\begin{figure}
   \centering
   \includegraphics[width=0.45\textwidth]{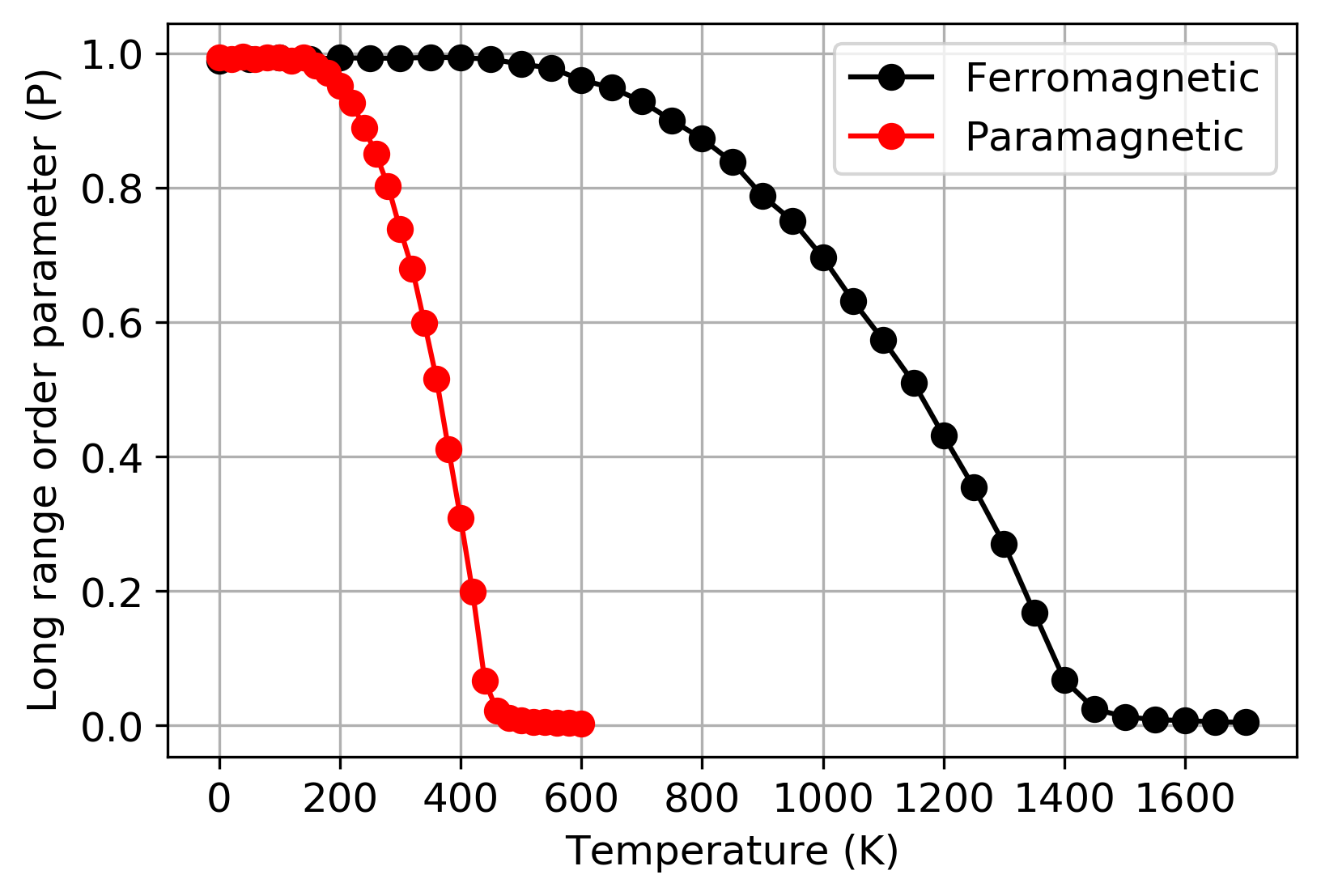}
   \caption{Long range order parameter \textit{P} as a function of temperature for the ferromagnetic (black) and paramagnetic (red) case, obtained from Monte Carlo simulations. 
   The temperature at which the long range order parameter vanishes, indicates the corresponding order-disorder transition temperature.}
   \label{img:p_vs_t}
\end{figure}

In order to estimate the order-disorder temperature from the calculated $E(P_z)$, we perform simple Monte Carlo simulations, as outlined in Sec.~\ref{sec:computational_details}.
The total energy for each Monte Carlo configuration is evaluated from the quadratic fit in Fig.~\ref{img:e_vs_p}, i.e., $E = - 4 N \Delta E P^2$, with $\Delta E = 62$\,meV. 
The resulting temperature dependence of the long range order parameter is shown in Fig.~\ref{img:p_vs_t} using a system size of $N=10^3$ (see Sec.~\ref{sec:computational_details}). Using larger system sizes does not lead to any noticeable changes.

It can be seen that the order parameter vanishes around 1400\,K, which is significantly higher than the reported experimental value for the order-disorder transition temperature of 593\,K~\cite{neel}. It is also significantly above the predicted ferromagnetic Curie temperature for L1$_0$ FeNi \cite{Alex,Lavrentiev2014}. Thus, assuming perfect ferromagnetic order when obtaining $\Delta E$ is probably not justified. 
In the following, we re-calculate the energy difference between chemically ordered and disordered states for the paramagnetic case, using the DLM approach~\cite{STAUNTON198415}, as described in Sec.~\ref{sec:DLM}).

\begin{figure}
   \centering
   \includegraphics[width=0.9\columnwidth]{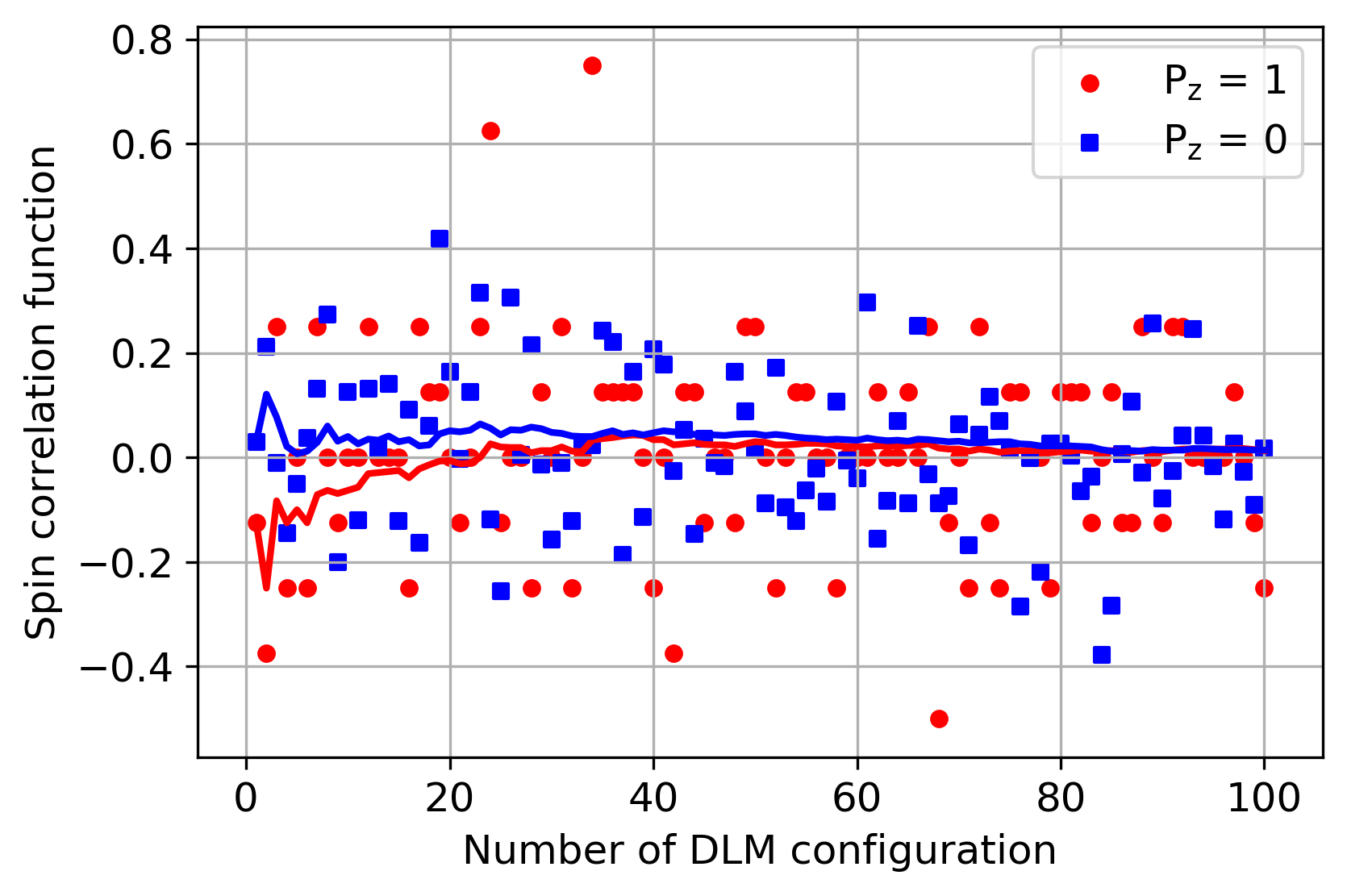}
   \caption{Calculated spin correlation functions for 100 DLM configurations for both chemically ordered ($P_z = 1$) and chemically disordered ($P_z = 0$) configurations. Solid red and blue lines represent the cumulative averages for $P_z = 1$ and $P_z = 0$, respectively.}
   \label{img:spin_correlation}
\end{figure}

To confirm that our sampling over a sufficient amount of randomly chosen DLM configurations converges as expected, Fig.~\ref{img:spin_correlation} shows the nearest-neighbour spin correlation function (see Eq.~\eqref{eq:spincorr}) for different chemically ordered and disordered magnetic configurations, evaluated from the converged magnetic moment directions, together with their cumulated averages, obtained by averaging over an increasing number of configurations.
One can see that the cumulated average of the spin correlation function approaches zero both for the chemically ordered and the chemically disordered magnetic configurations, which shows that the amount of configurations we average over is sufficient, and that the magnetic moments indeed converge to the directions that were initialized.

\begin{figure}
   \centering
   \includegraphics[width=\columnwidth]{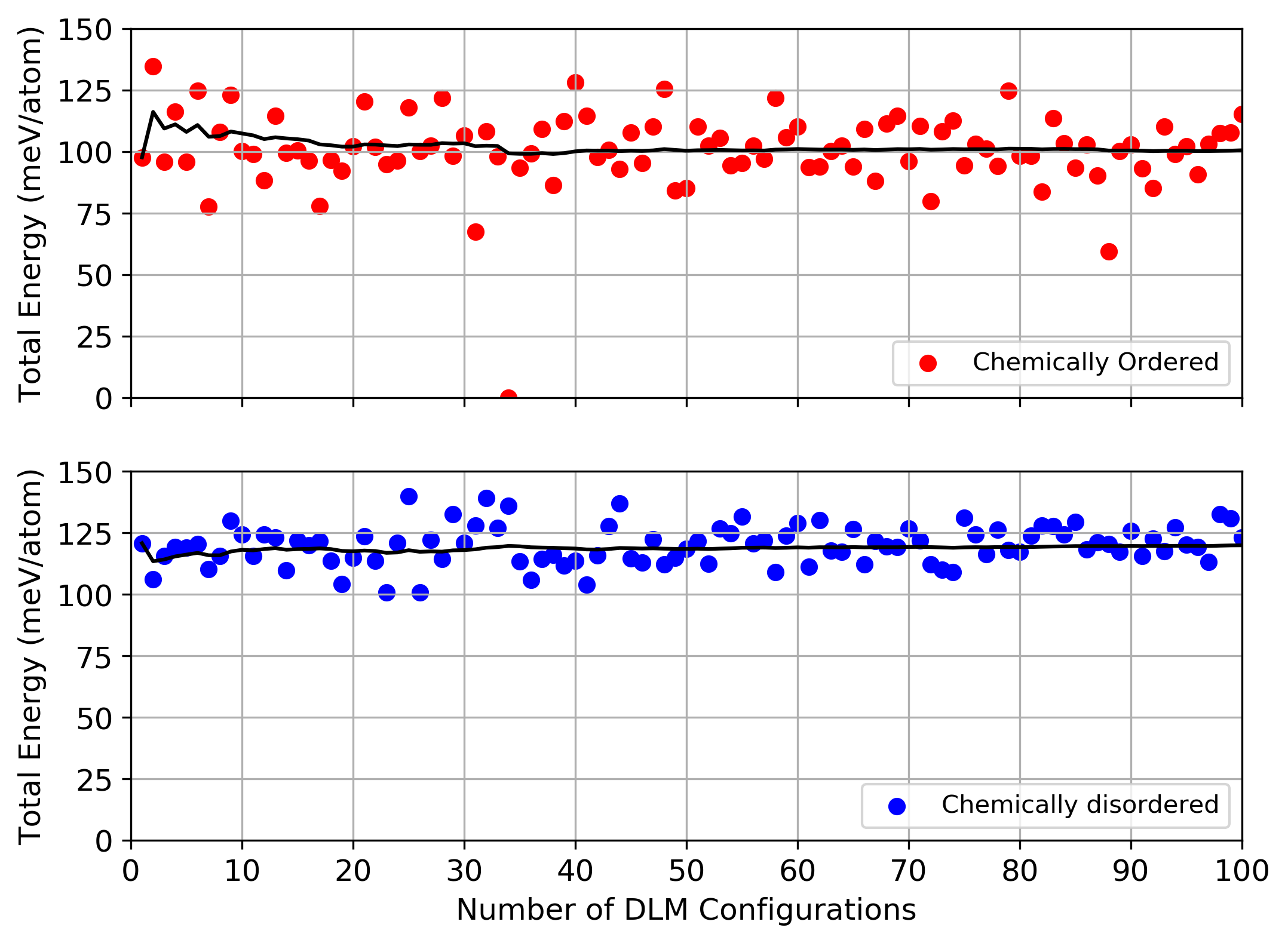}
   \caption{Calculated total energies (per atom) for 100 DLM configurations generated for the chemically ordered ($P_z=1$, top) and the chemically disordered ($P_z=0$, bottom) case, using a $2\times2\times2$ supercell. The cumulative averages are shown as solid black lines.}
   \label{img:energy_para}
\end{figure}

Fig.~\ref{img:energy_para} shows the total energies (per atom) obtained for the 100 DLM configurations corresponding to the chemically ordered ($P_z=1$) and the chemically disordered ($P_z=0$) case.
The cumulative averages are indicated by the solid black lines. All energies are taken relative to the chemically ordered ferromagnetic case. Again one can see that, in spite of the large variations in the energies of the individual configurations, the averages converge rather well, and appear to be accurate to a few meV already after averaging over about 50 configurations.

\begin{table}
\caption{\label{tab:table1}%
Average total energies (in meV/atom) of the ferromagnetic and the paramagnetic states for chemically ordered and chemically disordered FeNi (relative to the ferromagnetic chemically ordered case).}
\begin{ruledtabular}
\begin{tabular}{@{}lccc@{}}
 & L1$_0$-FeNi  & A1-FeNi\\
 & (chemically ordered) & (chemically disordered)\\
\hline
Ferromagnetic & 0  & 62 \\
Paramagnetic & 101 & 120  \\ 
\end{tabular}
\end{ruledtabular}
\end{table}

Table~\ref{tab:table1} summarizes the average total energies obtained for the ferromagnetic and paramagnetic state, both for the chemically ordered and the chemically disordered case.
It can be seen that the energy difference between the chemically ordered and the chemically disordered case is drastically reduced in the paramagnetic state compared to the ferromagnetic case (from 62\,meV to about 20\,meV per atom), indicating a strong coupling between chemical and magnetic order.
Furthermore, the energy difference between the ferromagnetic and the paramagnetic state is also significantly reduced in the chemically disordered alloy compared to the case with perfect L1$_0$ order (from about 100\,meV per atom to 58\,meV per atom).
This indicates that the magnetic Curie temperature of the chemically disordered phase is expected to be significantly lower than the (hypothetical) Curie temperature of the chemically ordered phase, which appears to be consistent with other theoretical studies \cite{Lavrentiev2014,Tian2019}.

For L1$_0$-ordered FeNi, a magnetic Curie temperature of $T_\text{C} = 916$\, K\, has been suggested, based on first principles DFT calculations \cite{Alex}. This is more or less consistent with the value of $\sim$ 1000\,K obtained from simulations using a first-principles-based Heisenberg-Landau magnetic cluster expansion~\cite{Lavrentiev2014}. 
However, on heating the L1$_0$ order starts to disappear at temperatures around 700-800\,K, depending somewhat on the heating rate~\cite{DOSSANTOS2015234}. Note that the actual reported chemical-order disorder temperature is much lower ($T_\text{od} = 593$\,K \cite{neel}), but that the chemical order is kinetically stable up to temperatures where atomic diffusion becomes thermally activated. Therefore, it is clear that the predicted $T_\text{C}$ for the ordered system is only a hypothetical Curie temperature, as the ordered phase is unstable at such high temperatures. 

If we simply scale the predicted values for $T_\text{C}$ of the chemically ordered case according to our obtained reduction of the ferromagnetic-paramagnetic energy difference, we obtain an estimate for the Curie temperature of chemically disordered FeNi of around 550\,K, which however appears too low compared to experimental values of around 785-789\,K~\cite{Onodera1981,Wei2014}. 

Interestingly, one should note that the temperature range where the chemical order effectively disappears ($\sim$ 700-800\,K~\cite{DOSSANTOS2015234}) is quite similar to the Curie temperature of the disordered system. This means that once the system disorders, the magnetic order also disappears rather abruptly (see, e.g., Refs.~\onlinecite{WASILEWSKI1988150,Lewis2014}). 

As seen in Fig.~\ref{img:p_vs_t}, the reduced $\Delta E$ obtained for the paramagnetic state also leads to a strong reduction of the order-disorder temperature, obtained in our simple Monte Carlo simulations, to about 450\,K. Note that these calculations are in principle expected to strongly underestimate the true order-disorder temperature, since the system is still magnetically ordered in that temperature range. On the other hand our simple approach neglects several other effects, e.g., lattice vibrations, which tend to reduce the order-disorder temperature \cite{vandeWalleRMP,MOHRI2009244,Tian2019}. Without considering such factors as well as kinetic effects, the temperature dependence of the order parameter is expected to follow the ferromagnetic curve for low temperatures and then move towards the paramagnetic curve once the magnetic order vanishes. Note, however, that the fully PM case considered here, with no short range correlations, is in principle only reached for $T \rightarrow \infty$.
Thus, while our simplified model is not expected to quantitatively predict the order-disorder transition temperature, it can provide order of magnitude estimates and clearly indicates the strong coupling between the chemical order-disorder transition and the magnetic state in L1$_0$-FeNi.

\subsection{\label{sec:MAE}Magneto-crystalline anisotropy}

Several studies in the past have investigated the MAE in L1$_0$ FeNi by means of first-principles calculations \cite{Alex, Miura2013, Lewis2014, freeman}. In addition, several experimentally measured values of MAE were also reported. However, very few investigations exists on the dependence of the MAE on the degree of chemical order in L1$_0$ FeNi. Kota and Sakuma~\cite{Kota2012} theoretically estimated the variation of MAE as a function of long-range order parameter for several L1$_0$ alloys including FeNi. They employed the tight-binding linear muffin-tin orbital method in conjunction with the CPA. They found that for FeNi, among other L1$_0$ alloys, the MAE is proportional to the power of the order parameter where the power varies from 1.6 to 2.4.

\begin{figure}
   \centering
   \includegraphics[width=0.45\textwidth]{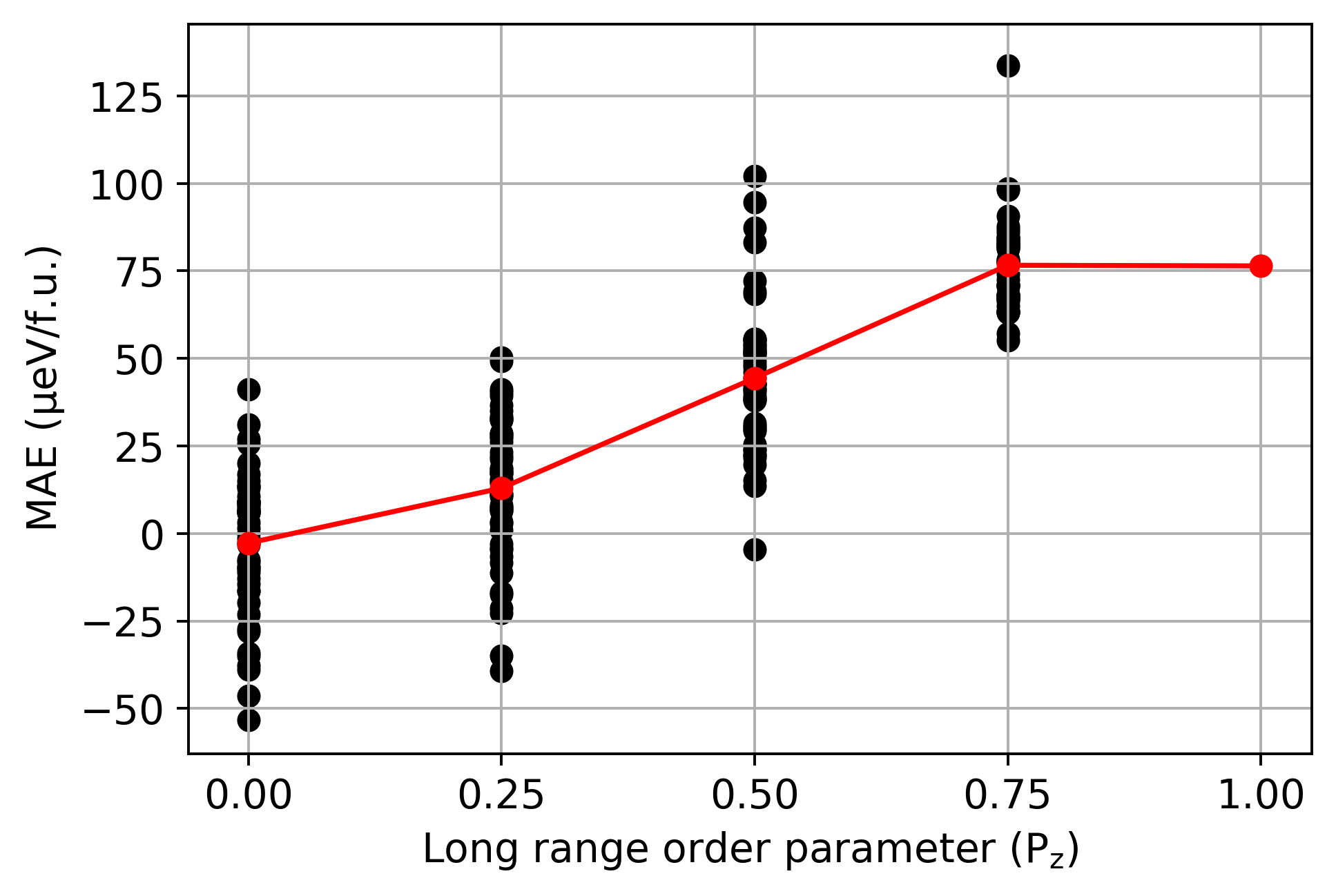}
   \caption{Calculated MAE, defined as $E_{[100]}-E_{[001]}$, as a function of the long range order parameter in FeNi, obtained for 50 different ferromagnetic configurations for each $P_z<1$. Black dots represent the  value for each configuration. Red dots correspond to the mean MAE for a particular $P_z$.}
   \label{img:MAE_vs_LRO}
\end{figure}

We calculate the dependence of the MAE on the long range order parameter by sampling over 50 ferromagnetic configurations for each value $P_z<1$, as described in Sec.~\ref{sec:computational_details}. Note that we also use the $2 \times 2 \times 2$ supercell to calculate the MAE for $P_z=1$ to obtain consistent data. The results are plotted in Fig.~\ref{img:MAE_vs_LRO}, which shows the data for each individual configuration as well as the average value for each $P_z$. 
It can be seen that for $P_z=0$, even though the MAE for the individual configurations shows a large spread of $\pm 50 \mu$eV/f.u., the obtained average is very close to the expected value of 0\,$\mu$eV/f.u. This indicates that we sample a sufficient amount of configurations to obtain reliable averages.

The MAE increases with increasing degree of chemical order, but, strikingly, reaches its maximal value already for $P_z=0.75$. This means that the MAE does not decrease significantly if the deviations from perfect order are not too large. In view of the fact that perfectly ordered samples are very difficult to synthesize, this is an important result.
We also note that our results do not follow the power-law behavior suggested by Kota and Sakuma ($\text{MAE}\propto P^{1.6\text{-}2.4}$)~\cite{Kota2012}. This is most likely due to their use of the CPA approximation to describe compositional disorder and shows that for a quantity such as the MAE, effects of the local environment can be very important. This is different from the total energy, shown in Fig.~\ref{img:e_vs_p}, which agrees well with previous CPA calculations~\cite{Tian2019}.
Furthermore, for both $P_z = 0.75$ and $P_z=0.5$, we find some configurations with even higher MAE than the fully ordered alloy. This indicates, that it might be possible to further increase the anisotropy of this system,  beyond the value obtained for the stoichiometric 50:50 composition with perfect chemical order.

The MAE we obtain for the fully ordered case ($P_z=1$) is 76\,$\mu$eV/f.u. corresponding to 0.54\,MJ/m$^3$, which agrees well with previous calculations using similar methods (0.56\,MJ/m$^3$~\cite{Miura2013}, 0.48\,MJ/m$^3$~\cite{Alex}, and 0.47\,MJ/m$^3$~\cite{Werwi_ski_2017}). 
We note that this value is quite comparable, albeit slightly smaller, than what has been reported experimentally in Ref.~\cite{Kojima_2011} for samples with a long-range order parameter around 0.5 ($\approx 0.7$\,MJ/m$^3$).
On the other hand, for $P_z=0.5$, we obtain a value that is clearly smaller than the experimentally reported MAE. This suggests that we are underestimating the true MAE of the system. Indeed, it has been shown, that including a so-called \emph{orbital polarization correction} can enhance the MAE of the fully ordered system roughly by a factor of two~\cite{Eriksson2001, Miura2013}.

\subsection{Orbital magnetic moment anisotropy}

In order to obtain further insights into the origin of the MAE, we now analyze the orbital magnetic moment anisotropy as a function of long range order parameter. 
The orbital magnetic moment and its anisotropy is often closely connected to the MAE~\cite{Bruno1989,vanderLaan1998}. In the present case it can potentially provide insights as to which local chemical environments are particularly favorable for obtaining a large MAE.
We define the orbital moment anisotropy as $\Delta L = L_{[001]} - L_{[100]}$, where $L_{[001]}$ and $L_{[100]}$ are the total orbital magnetic moments (summed over all atoms in the $2 \times 2 \times 2$ supercell) when the magnetization lies along the $[001]$ and $[100]$ directions, respectively. Here, the sign is chosen such that the orbital anisotropy is positive if the orbital magnetic moments are larger along the $[001]$ direction (which is the easy magnetic axis for L1$_0$ FeNi).

\begin{figure}
   \centering
   \includegraphics[width=0.45\textwidth]{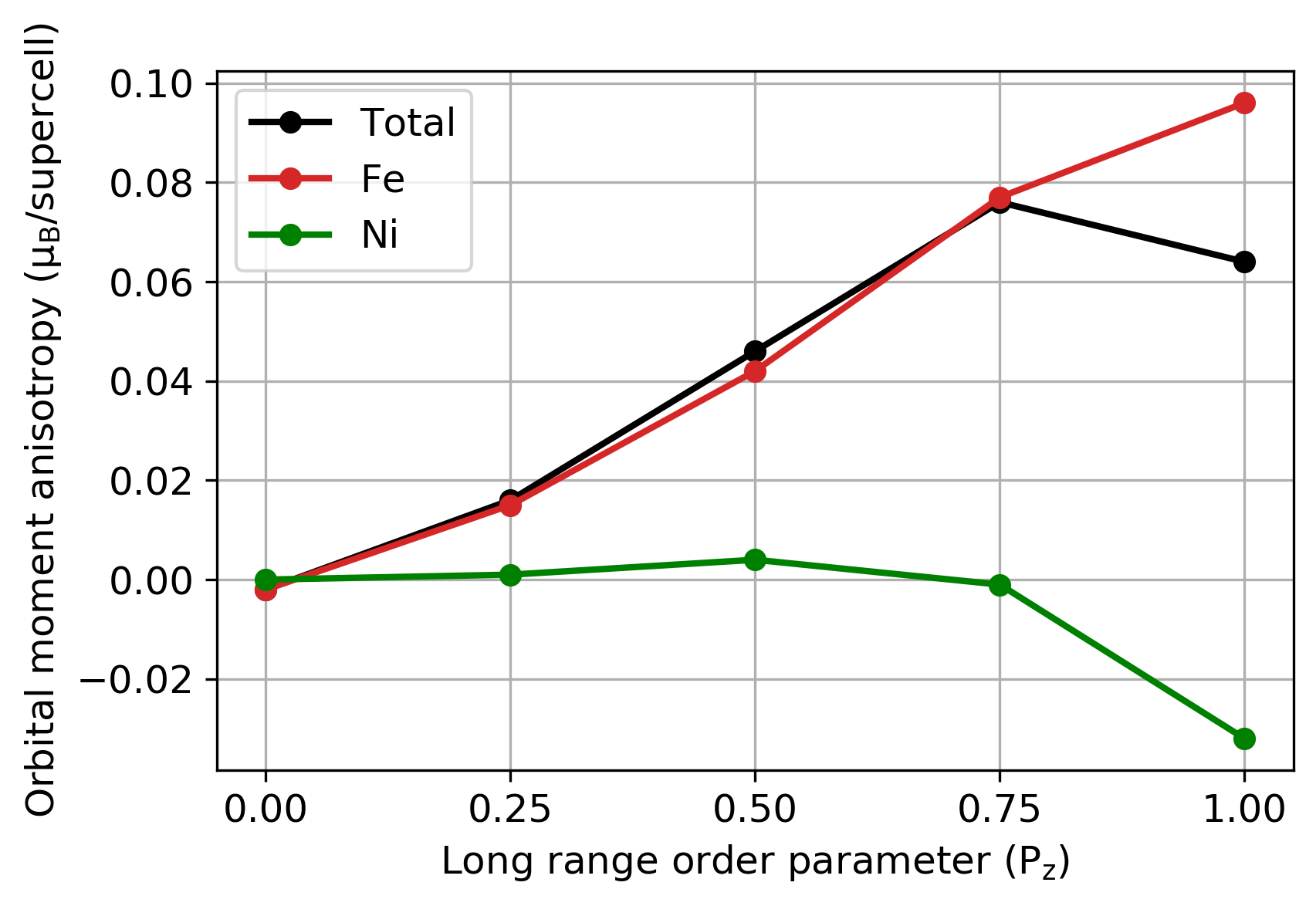}
   \caption{Calculated total orbital moment anisotropy, $\Delta L = L_{[001]} - L_{[100]}$, averaged over all configurations with the same long range order parameter $P_z$, as a function of $P_z$ Separate contributions of all Fe and all Ni atoms in the system are also shown.}
   \label{img:oma_vs_lro}
\end{figure}

Fig.~\ref{img:oma_vs_lro} shows the total as well as the atom-resolved orbital moment anisotropy as a function of the long range order parameter (i.e., averaged over all configurations corresponding to the same $P_z$). One can clearly see that the main contribution to the total orbital moment anisotropy for $P_z < 1$ comes from the anisotropy of the Fe orbital magnetic moment, while the contribution from the Ni moments is almost negligible. For the perfectly ordered structure, we observe that the orbital magnetic moments of the Ni atoms are larger along the $[100]$ direction, which results in a small decrease of the total orbital moment anisotropy as we go from $P_z = 0.75$ to $P_z = 1$ (see solid black curve in the Fig \ref{img:oma_vs_lro}). 

Note that both the MAE and the total orbital moment anisotropy show similar behaviour as one increases the long range order in the system. 
This suggests a possible explanation for the somewhat unexpected behavior of the MAE, provided that the MAE can be understood in terms of local contributions of the Fe and Ni atoms that correlate with the corresponding orbital moment anisotropies. 
Thereby, the (small) contribution to the MAE from the Ni atoms would be opposite to that of the Fe atoms and also be much more sensitive to deviations from perfect chemical order, such that it essentially vanishes already for $P_z \leq 0.75$, while the contribution from the Fe is still rather strong.

\begin{figure}
   \centering
   \includegraphics[width=0.45\textwidth]{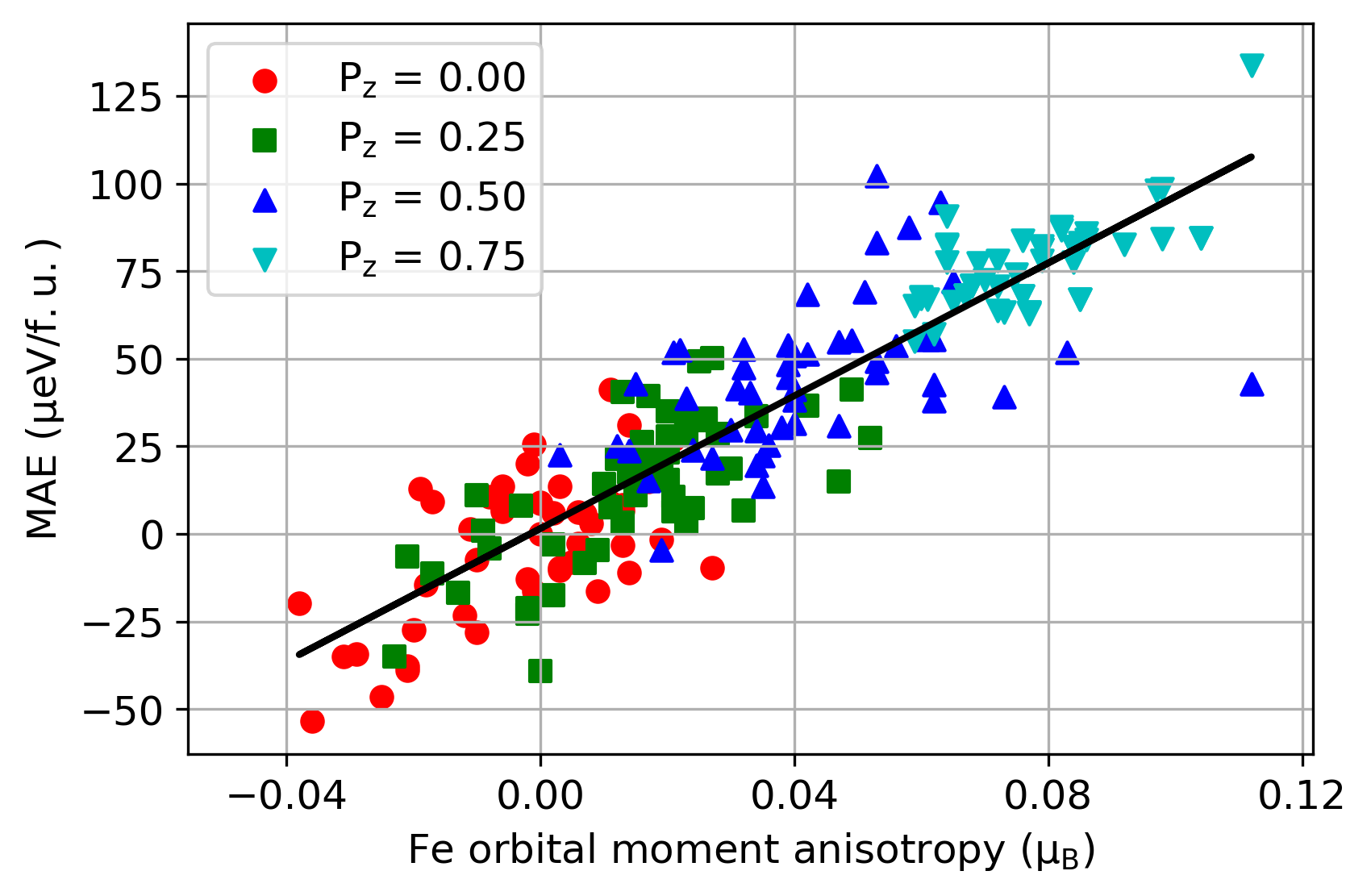}
   \caption{MAE versus orbital moment anisotropy of the Fe atoms for each individual configuration. Configurations corresponding to different values of $P_z$ are indicated by different markers. The solid black line corresponds to a least square fit to the data.}
   \label{img:MAE_vs_oma}
\end{figure}

To further demonstrate the correlation between the MAE and the orbital moment anisotropy, we show in Fig.~\ref{img:MAE_vs_oma}, the MAE as a function of orbital moment anisotropy for all individual configurations with different values of the chemical order parameter. Only the orbital moment anisotropy obtained from the Fe atoms is shown, here. There is a clear linear correlation between the two quantities, indicated also by the least mean square fit to all data points (solid black line). 
On the other hand, there can also be a noticeable spread in the linear relationship between the MAE and the orbital moment anisotropy on the level of the individual configurations.
Nevertheless, our results suggest that the orbital moment anisotropy can in principle be used as proxy for the MAE, which then allows to analyze how the local environment affects both quantities. 

\section{\label{sec:Summary}Summary and Conclusions}

In summary, we have demonstrated a strong coupling between chemical and magnetic orders in L1$_0$ FeNi, consistent with previous studies employing effective medium/mean-field type approaches to describe the compositional disorder. Specifically, our results show that chemical disorder reduces the energy difference between the ferromagnetic and paramagnetic state by about 40\,\%.
Consequently, the magnetic Curie temperature of the disordered system is much lower than the rather high (hypothetical) Curie temperature of the ordered phase. As a result, the magnetic order vanishes once the system starts to disorder under heating, as has been observed in various experiments~\cite{WASILEWSKI1988150,Lewis2014}.
On the other hand, perfect ferromagnetic order increases the energy gain due to chemical order by nearly a factor of three compared to the paramagnetic case. In principle, this implies, that if it would somehow be possible to stabilize the ferromagnetic state at higher temperatures, one could artificially increase the order-disorder transition temperature, which could then ease the synthesis of the ordered material.
While our simple energetic model is obviously too crude to obtain very accurate values for the order-disorder transition temperature, the estimates we obtain from our Monte Carlo simulations give the correct order of magnitude, indicating that our DFT calculations correctly describe the underlying energetics.

Most importantly, our calculations of the magneto-crystalline anisotropy (MAE) as function of the chemical long-range order parameter $P_z$ reveal that a reduction of $P_z$ by 25\,\% does not decrease the MAE within the accuracy of our method. This is rather encouraging, since it shows that full chemical order is not required to obtain full anisotropy. However, it also indicates that previous estimates of the full anisotropy, based on the extrapolation of results  obtained for partially ordered samples, are probably too high.
We note that in order to obtain this result, the use of our  configurational sampling method is crucial. Effective medium approaches, such as CPA, do not take into account the specific local chemical environment and thus will always predict a gradual decrease of the MAE for reduced chemical order.

Interestingly, we obtain the highest MAE for certain configurations with partial disorder, which suggests that the MAE can potentially be increased beyond the value obtained for the perfectly ordered L1$_0$ structure. 
We also demonstrate a clear correlation between the orbital magnetic moment anisotropy and the MAE, which suggests that chemical environments resulting in a large local orbital moment anisotropy will also be favorable for obtaining a high MAE. 
While it might be highly non-trivial to engineer a specific partially disordered configuration, it provides an exciting avenue to optimize the MAE in tetrataenite with respect to the local chemical environment, by considering small deviations from perfect L1$_0$ order as well as from the ideal equiatomic stoichiometry.
 
\begin{acknowledgments}
This work was supported by ETH Z\"urich. Calculations were performed on the cluster \enquote{Piz Daint}, hosted by the Swiss National Supercomputing Centre, and the \enquote{Euler} cluster of ETH Z\"urich.
\end{acknowledgments}

\bibliography{main}
\end{document}